\documentclass[fleqn,usenatbib,twocolumn]{mnras}
\usepackage{pdflscape}
\usepackage{newtxtext,newtxmath}
\usepackage{multirow}

\usepackage[T1]{fontenc}

\DeclareRobustCommand{\VAN}[3]{#2}
\let\VANthebibliography\thebibliography
\def\thebibliography{\DeclareRobustCommand{\VAN}[3]{##3}\VANthebibliography}

\usepackage{graphicx}	
\usepackage{amsmath}	

\title[The temporally evolving statistics of two FRBs]{The temporally evolving energy and waiting time statistics of two repeating fast radio bursts}

\author[Y. Sang and H.-N. Lin]{
Yu Sang$^{1, 2}$,
Hai-Nan Lin$^{3, 4}$\thanks{Corresponding author: linhn@cqu.edu.cn}\\
$^{1}$Center for Gravitation and Cosmology, College of Physical Science and Technology, Yangzhou University, Yangzhou 225009, China\\
$^{2}$Shanghai Frontier Science Research Center for Gravitational Wave Detection, School of Aeronautics and Astronautics,\\ Shanghai Jiao Tong University, Shanghai 200240, China\\
$^{3}$Department of Physics, Chongqing University, Chongqing 401331, China\\
$^{4}$Chongqing Key Laboratory for Strongly Coupled Physics, Chongqing University, Chongqing 401331, China
}

\date{Accepted XXX. Received YYY; in original form ZZZ}

\pubyear{2023}

\begin{document}
\label{firstpage}
\pagerange{\pageref{firstpage}--\pageref{lastpage}}
\maketitle

\begin{abstract}
Based on two very large samples of repeating fast radio bursts (FRBs), i.e. FRB 20121102A and FRB 20201124A observed by the FAST telescope, we study the statistical properties of energy and waiting time. The bent power-law (BPL) model, thresholded power-law (TPL) model and Band function are used to fit the distribution of energy, and the BPL model and exponential (EXP) model are used to fit the distribution of waiting time. It is found that no single model can fit the distribution of energy or waiting time well in the full range. To investigate the possible temporal evolution, we divide the full samples into several subsamples according to the  observing sessions. We find that the distribution of energy for all subsamples can be well fitted by both BPL model and TPL model, while the distribution of waiting time for all subsamples can be well fitted by both BPL model and EXP model. Importantly, for the distribution of energy, the BPL index $\beta$ of all the subsamples is almost invariant, but the median value parameter $x_b$ varies significantly. Similar situation happens in the distribution of waiting time. Furthermore, for the distribution of waiting time, the occurrence rate parameter $\lambda$ in EXP model varies significantly. These features show that there may be a common emission mechanism for repeating FRBs, but the burst energy and occurrence rate are temporally evolving.

\end{abstract}

\begin{keywords}
 fast radio bursts -- methods: statistical -- radio continuum: transients
\end{keywords}



\section{Introduction}

Fast radio bursts (FRBs) are radio transients with millisecond-duration and extreme brightness temperatures \citep{Cordes:2019cmq,Petroff:2019tty,Platts:2018hiy,Xiao:2021omr,Zhang:2020qgp,2022arXiv221203972Z}. 
Since the first discovery in 2007 \citep{Lorimer:2007qn}, hundreds of FRBs have been detected \citep{Petroff:2016tcr,CHIMEFRB:2021srp}. FRBs are usually classified into repeaters and non-repeaters according to the number of detected bursts from the same source. Although tens of repeaters have been observed, most of them are not very active \citep{CHIMEFRB:2021srp}. There are two exceptions, FRB 20121102A and FRB 20201124A, from which thousands of bursts have been detected \citep{Li:2021hpl,Xu:2021qdn,Zhou:2022nnh}.

FRB 20121102A is the first detected repeater and has been observed from 400 MHz to 8 GHz \citep{Spitler:2014fla,Spitler:2016dmz,Chatterjee:2017dqg,Gajjar:2018bth,Michilli:2018zec}. 
First discovered in the Arecibo Pulsar ALFA Survey at 1.4 GHz \citep{Spitler:2014fla}, FRB 20121102A was found to be repeating by observations of 10 additional bursts with dispersion measures and sky positions consistent with the first burst \citep{Spitler:2016dmz}. Later on, its host galaxy was identified as a low-metallicity star-forming dwarf galaxy at cosmological distance ($z = 0.19$) \citep{Bassa:2017tke,Chatterjee:2017dqg,Marcote:2017wan,Tendulkar:2017vuq}.
Prior to detection by Five-hundred-meter Aperture Spherical radio Telescope (FAST), only $\sim 300$ bursts had been observed from FRB 20121102A \citep{Petroff:2016tcr,Zhang:2018jux,Gourdji:2019lht}. Recently, FAST significantly increased this sample with the detection of 1652 additional bursts \citep{Li:2021hpl}.

FRB 20201124A was first detected on 2020 November 24 by the Canadian Hydrogen Intensity Mapping Experiment (CHIME), and about four minutes later it was observed to repeat \citep{2021ATel14497....1C,Lanman:2021yba}. 
It entered a period of high activity in 2021 March and repeating bursts were detected by several observatories \citep{2021ATel14502....1K,2021ATel14508....1K,2021ATel14518....1X}. 
The localization was obtained to arcsecond precision and the host galaxy was identified by the Australia Square Kilometre Array Pathfinder (ASKAP) \citep{2021ATel14515....1D}. Later on, the spectroscopic redshift of the host galaxy was measured to be  $z = 0.0979 \pm 0.0001$ by the MMT Observatory \citep{Fong:2021xxj}. A large sample of 1863 bursts were detected by FAST from 2021 April 1 to June 11 at 1.0 -- 1.5 GHz \citep{Xu:2021qdn}. Several months later, FAST detected that FRB 20201124A entered an extremely active episode in the end of September 2021 \citep{Zhou:2022nnh,Zhang:2022rib}.

The statistical properties of the repeating FRBs are helpful to understand the emission mechanism of FRBs. Several works have studied the energy and waiting time (defined as the detected time interval of two adjacent bursts) distribution of FRBs \citep{Wang:2016lhy,Wang:2019sio,Lin:2019ldn,Zhang:2021ztz,Wang:2022gmu,Lu:2016fgg,Li:2016qbl,Macquart:2018jlq,Wang:2017agh,Lu:2019pdn,Wang:2019suh}. Based on 17 bursts from the repeating FRB 20121102A, the distributions of fluence, peak flux, duration and waiting time were found to follow the power-law distribution \citep{Wang:2016lhy}.
A universal energy distribution with power-law form was found for FRB 20121102A in six samples observed by different telescopes at different frequencies, and the power-law index is similar to the non-repeating FRBs \citep{Wang:2019sio}.
Based on two samples including 93 and 41 bursts from FRB 20121102A observed by Green Bank Telescope and Arecibo Observatory, the cumulative distributions of fluence, flux, energy and waiting time were found to be consistent with the bent power law \citep{Lin:2019ldn}. Moreover, the probability density functions of fluctuations of fluence, flux and energy were found to follow a scale-invariant Tsallis $q$-Gaussian distribution \citep{Lin:2019ldn}. The statistical properties of the power-law energy distribution and scale-invariant Tsallis $q$-Gaussian fluctuations were also found in earthquakes and soft gamma repeaters (SGRs), implying that there may be some similar occurrence mechanism between these events, which could be explained by the self-organized criticality processes \citep{Wang:2015nsl,Wang:2016lhy,Chang:2017bnb,Wang:2017agh,Lin:2019ldn,Cheng:2019ykn,Wei:2021kdw,Sang:2021cjq,Wang:2022gmu}.

In this paper, we further study the statistical properties of two repeating FRBs, i.e. FRB 20121102A and FRB 20201124A, based on two large samples observed by FAST \citep{Li:2021hpl,Xu:2021qdn}. The statistics of these two samples have been investigated in the original papers \citep{Li:2021hpl,Xu:2021qdn}. However, we note that the bursts are detected in a long observational period spanning several months, with some observational gaps. Therefore, we focus on the temporally evolving statistical properties of these two FRBs, which has not been done before. The rest of the paper is arranged as follows: The samples and models are introduced in Section \ref{sec:data}. In Section \ref{sec:full-sample}, we study the statistical properties of the full samples. In Section \ref{sec:sub-sample}, we study the temporal evolution of statistical properties by dividing the full samples into several observing sessions. Finally, discussions and conclusions are given in Section \ref{sec:conclusion}.

\section{Data and models}\label{sec:data}

Our analysis is based on two samples observed by FAST. The first sample includes 1652 bursts from the repeating FRB 20121102A detected in a total of 59.5 observing hours between August 29 and October 29, 2019 \citep{Li:2021hpl}. This sample reduces the flux lower limit by at least three times than previously published bursts from this source. The derived isotropic equivalent energy at 1.25 GHz spans more than three orders of magnitude. The bimodal energy distribution has a prominent peak and two broad bumps, which can not be fitted by a single power-law or log-normal function, but can be well fitted by a log-normal distribution plus a generalized Cauchy function \citep{Li:2021hpl}. Some parts of the distribution are consistent with a simple power law (at high energies), the log-normal distribution (at low energy end), or the generalized Cauchy function (at high energy end). The waiting times are found to be well consistent with two log-normal distribution for the full sample, with the main peak at 70 s and the secondary peak at 3.4 ms. After energy cut $E > 3 \times 10 ^{38} {\rm erg}$, the waiting time distribution can be well fitted by a log-normal function peaking at 220 s.

The second sample includes 1863 bursts from the repeating FRB 20201124A detected from April 1 to June 11, 2021, with a total of 82 hours observation time in the frequency range of 1.0 -- 1.5 GHz \citep{Xu:2021qdn}.
A broken power law can fit the cumulative distribution function (CDF) of the burst energy better than a single power law. The best-fitting broken power law is connected by two simple power law with indices $0.36$ and $1.5$ at the broken point $E_0= 1.1 \times 10 ^{38} {\rm erg}$. The power-law index at high energy end is close to those of FRB 20121102A and FRB 20180916B. At low energy end the power-law index is shallower than that of FRB 20121102A. The distribution of waiting time can be fitted by the superposition of three log-normal functions, with peaks at 39 ms, 45.1 s and 162.3 s, respectively. The similar multimodal distribution of waiting times are also found in FRB 20121102A \citep{Li:2021hpl}.

In this paper, we further study the distributions of energy and waiting time using the samples of FRB 20121102A and FRB 20201124A, respectively. Following \citet{Xu:2021qdn}, the isotropic burst energy is calculated by $E = 4\pi D_L^2 (1+z)^{-1} \int F d BW$, where F is the observed fluence, BW is the signal bandwidth, $z$ is the redshift, and $D_L$ is the luminosity distance. The redshifts of FRB 20121102A and FRB 20201124A are $z = 0.193$ and $z = 0.09795$, respectively, and the corresponding luminosity distances are $D_L = 949$ Mpc and $D_L = 453.3$ Mpc, respectively. We ignore the waiting times which are larger than 1 hr, to avoid the large observational gaps. We use three models to fit the CDFs of energy and waiting time.

The first model used to fit the CDF of energy is the bent power law (BPL) distribution, which is written as \citep{Guidorzi:2016ddt}
\begin{equation}
  N(>x)=B\left[1+\left(\frac{x}{x_b}\right)^{\beta}\right]^{-1},
\end{equation}
where $B$ is the total number of bursts, $x_b$ is the median value of $x$, and $\beta$ is the power-law index. Here $x$ is the quantity that we are interested in, either the energy ($E$), or the waiting time (WT). The BPL function is a piecewise power law smoothly connected, which has a very shallow power-law index at left end ($x\ll x_b$), and reduces to simple power law $x^{-\beta}$ at right end ($x\gg x_b$). Note that the BPL model is a phenomenological model, but it is found to be consistent with the energy and waiting time distributions of SGRs \citep{Chang:2017bnb,Sang:2021cjq} and repeating FRBs \citep{Lin:2019ldn,Wang:2022gmu}.

The thresholded power-law (TPL) model was derived and first used to fit the distribution of solar flares and stellar flares in \citet{Aschwanden:2015}. Later on, it was shown that the TPL model can well fit the energy distribution of FRBs and SGRs \citep{Cheng:2019ykn,Sang:2021cjq}. The differential TPL distribution is given by \citep{Aschwanden:2015}
\begin{equation}
   N(x)dx = n_0 (x_0 + x)^{- \gamma } dx,\quad x_1\leq x\leq x_2,
\end{equation}
where $n_0$ is the normalization constant, $x_0$ is the threshold parameter. $x_1$ and $x_2$ are the minimum and maximum values of the data points, respectively. The cumulative distribution of the TPL model is given by the integral of the differential TPL function,
\begin{equation}
   N( >x) =  \int_{x}^{x_2} N(x)dx  = \frac{n_0}{1-  \gamma } [(x_0 + x_2)^{1- \gamma } - (x_0 + x)^{1- \gamma } ].
\end{equation}
Here $x_0$, $n_0$ and $\gamma$ are free model parameters.

The Band function is an exponentially connected broken-power-law function first proposed to describe the spectrum of gamma-ray bursts, which reads \citep{Band:1993eg}
\begin{equation}
    N(>x)=\left\{\begin{aligned}
    & Ax^{\hat\alpha}\, {\rm e}^{\left(-x/x_c\right)} \quad &x\le(\hat\alpha-\hat\beta)x_c,\\
       & Ax^{\hat\beta} \, \left[\frac{(\hat\alpha-\hat\beta)x_c}{\rm e}\right]^{\hat\alpha-\hat\beta} \quad &x\ge(\hat\alpha-\hat\beta)x_c.\\
    \end{aligned}\right.
\end{equation}
Here $\hat\alpha$ and $\hat\beta$ are the power-law indices of the lower and higher energy part of the distribution respectively, $x_c$ is a characteristic value for $x$, and $A$ is the normalization factor. The Band function was found to fit the CDF of energy of FRB 20201124A reasonably well \citep{Zhang:2022rib}.

If the emission of bursts is a simple Poisson process, and the bursts randomly and independently occur in time, then the waiting time is expected to follow the exponential (EXP) distribution with a constant average occurrence rate $\lambda$. In this case, the CDF of waiting time is given by
\begin{equation}
   N(>\Delta t) \propto e^{-\lambda \Delta t}.
\end{equation}
The EXP model was used to fit the waiting time distribution of SGRs \citep{Sang:2021cjq} and repeating FRBs \citep{Wang:2017agh,Lin:2019ldn}.

In the next two sections, we first use the full samples to constrain the model parameters (Section \ref{sec:full-sample}), then we consider the possible temporal evolution by dividing the full samples into several subsamples according to the observing sessions (Section \ref{sec:sub-sample}). The model parameters are constrained by minimizing the $\chi^2$,
\begin{equation}
  \chi^2=\sum_i\frac{[N_i-N(>x_i)]^2}{\sigma_i^2},
\end{equation}
where the uncertainty of the data point is taken to be $\sigma_i=\sqrt{N_i}$ in the calculations.

\section{The statistics of the full sample}\label{sec:full-sample}

Based on the full samples of FRB 20121102A and FRB 20201124A, we use the BPL, TPL and Band models to fit the CDF of energy, and use the BPL and EXP models to fit the CDF of waiting time.

\begin{figure*}
    \centering
	\includegraphics[width=0.48\textwidth]{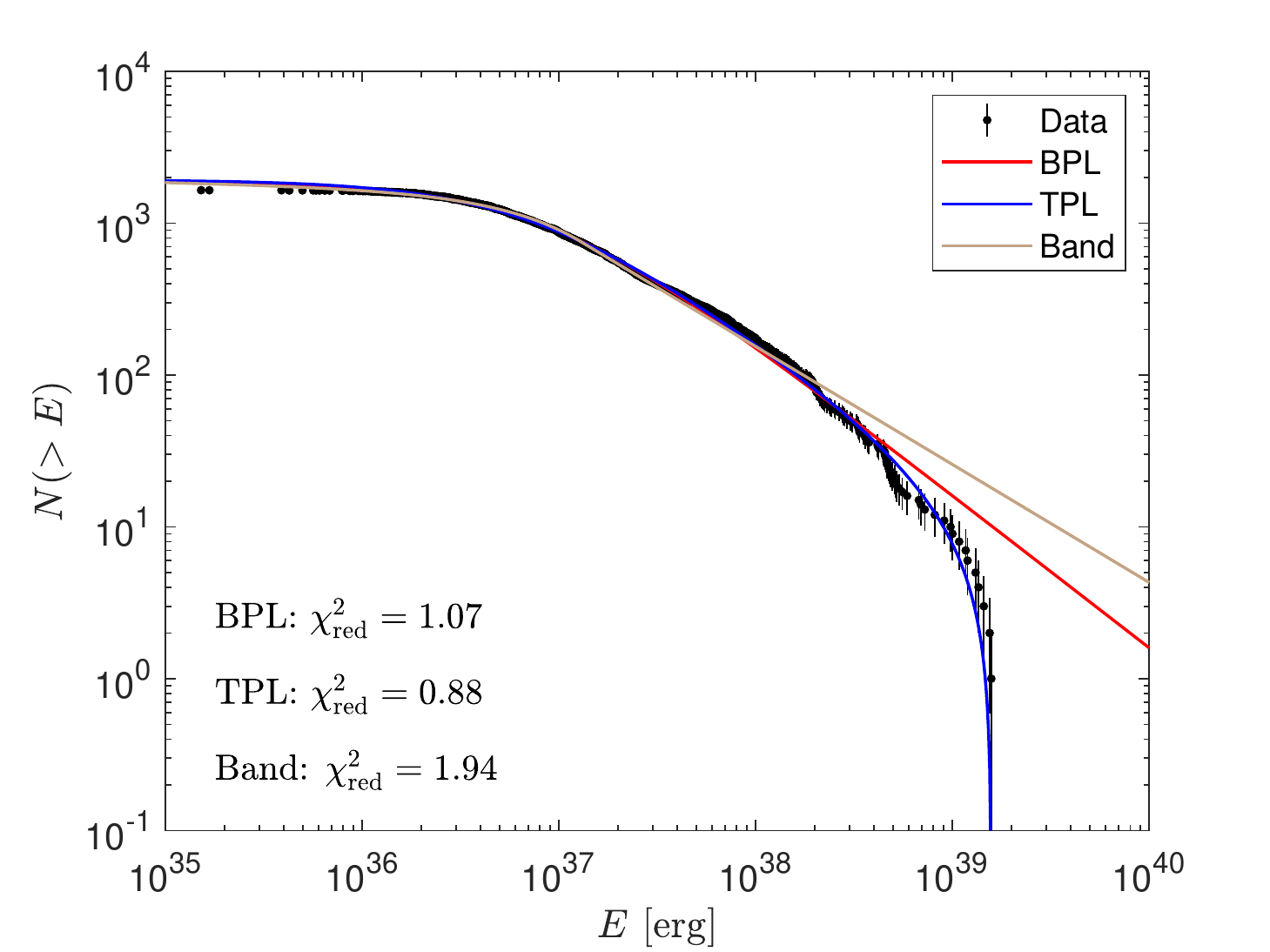}
	\includegraphics[width=0.48\textwidth]{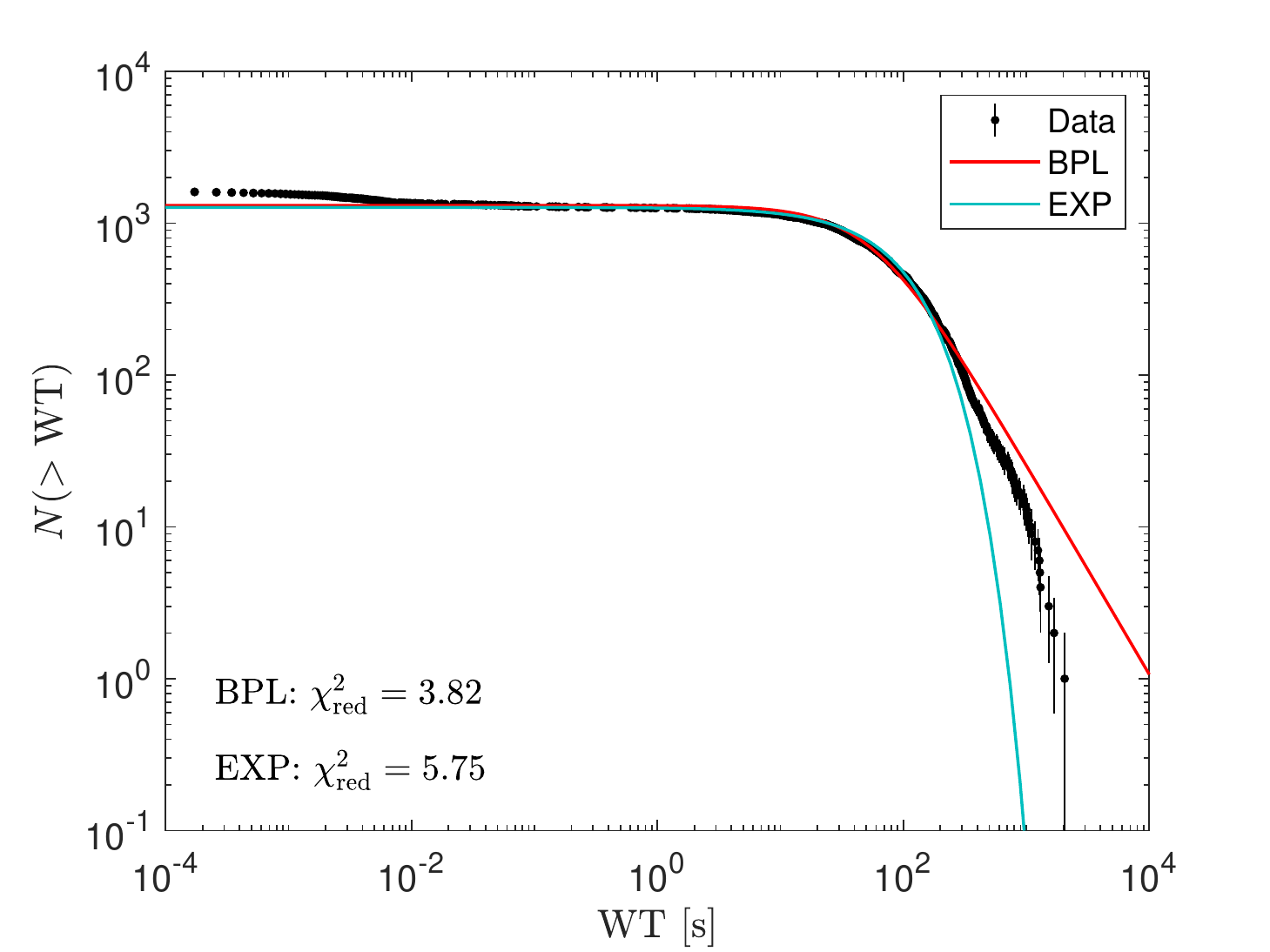}
    \caption{The CDFs of energy (left panel) and waiting time (right panel) for FRB 20121102A. In the left panel, the red, blue and light brown solid lines are the best-fitting curves to the BPL, TPL and Band models, respectively. In the right panel, the red and green solid lines are the best-fitting curves to the BPL and EXP models, respectively.}
    \label{fig:frb121102_full}
\end{figure*}

The left panel of Figure \ref{fig:frb121102_full} shows the CDF of energy for FRB 20121102A, and the corresponding model parameters are summarized in Table \ref{tab:frb121102_full}. The BPL model fits the data points at left and middle parts very well, but fails at high energy end ($E\gtrsim 5\times10^{38} {\rm erg}$). The best-fitting parameters for BPL model are $\beta = 1.00\pm0.01$ and $x_b = (0.87\pm0.01)\times 10^{37} {\rm erg}$. The TPL model is consistent with the data points in most energy range, except for the very left-most end and a few points at $E\sim 6\times 10^{38} {\rm erg}$. The best-fitting parameters for TPL model are $\gamma = 1.90\pm0.01$ and $x_0 = (0.72\pm0.01) \times 10^{37} {\rm erg}$. TPL model fits the CDF of energy better than BPL model, as the reduced chi-square value of TPL model ($\chi^2_{\rm red}= 0.88$) is smaller than that of BPL model ($\chi^2_{\rm red}= 1.07$). The best-fitting parameters for Band model are $\hat\alpha=-0.03\pm0.01$, $\hat\beta=-0.77\pm0.01$ and $x_c=(1.71\pm 0.02)\times{\rm 10^{37}erg}$, with the reduced chi-square value $\chi^2_{\rm red}=1.94$. The Band model fits the data much worse than both BPL model and TPL model. More specifically, the Band model fails to fit the data at the right-most end.

The right panel of Figure \ref{fig:frb121102_full} shows the CDF of waiting time for FRB 20121102A, and the corresponding model parameters are summarized in Table \ref{tab:frb121102_full}. The waiting time values larger than 1 hour are ignored to avoid long observational gaps. The best-fitting BPL parameters are $\beta = 1.38\pm0.01$ and $x_b = 58.65\pm0.41 {\rm s}$, and the best-fitting EXP parameters are $\lambda = (9.79\pm0.05) \times 10^{-3} {\rm s}^{-1}$. As we can see, both BPL and EXP models fit the data well at middle parts, but deviate from the data points at left and right ends. At left end, both models are below the data points. At right end, the BPL prediction is above the data points, while the EXP prediction is below the data points. In a whole, both models fail to fit the waiting time distribution, since the reduced chi-square values of BPL model ($\chi^2_{\rm red}= 3.82$) and EXP model ($\chi^2_{\rm red}= 5.75$) are very large. 

\begin{table*}
\centering
\caption{The best-fitting parameters of the BPL, TPL, Band and EXP models for FRB 20121102A.}\label{tab:frb121102_full}

{\begin{tabular}{lllll} 
\hline\hline 
\multicolumn{5}{c}{energy}\\
\hline
BPL & & $\beta=1.00\pm0.01$  & $x_b=(0.87\pm0.01)\times{\rm 10^{37}erg}$ & $\chi^2_{\rm red}=1.07$\\
TPL & & $\gamma=1.90\pm0.01$ & $x_0=(0.72\pm0.01)\times{\rm 10^{37}erg}$ & $\chi^2_{\rm red}=0.88$\\
Band & $\hat\alpha=-0.03\pm0.01$ & $\hat\beta=-0.77\pm0.01$ & $x_c=(1.71\pm 0.02)\times{\rm 10^{37}erg}$ & $\chi^2_{\rm red}=1.94$\\
\hline
\multicolumn{5}{c}{waiting time}\\
\hline
BPL & & $\beta=1.38\pm0.01$ & $x_b=58.65\pm0.41~{\rm s}$ & $\chi^2_{\rm red}=3.82$\\
EXP & & & $\lambda=(9.79\pm0.05)\times{\rm 10^{-3}s^{-1}}$ & $\chi^2_{\rm red}=5.75$\\
\hline 
\end{tabular}}  
\end{table*}

\begin{figure*}
    \centering
	\includegraphics[width=0.48\textwidth]{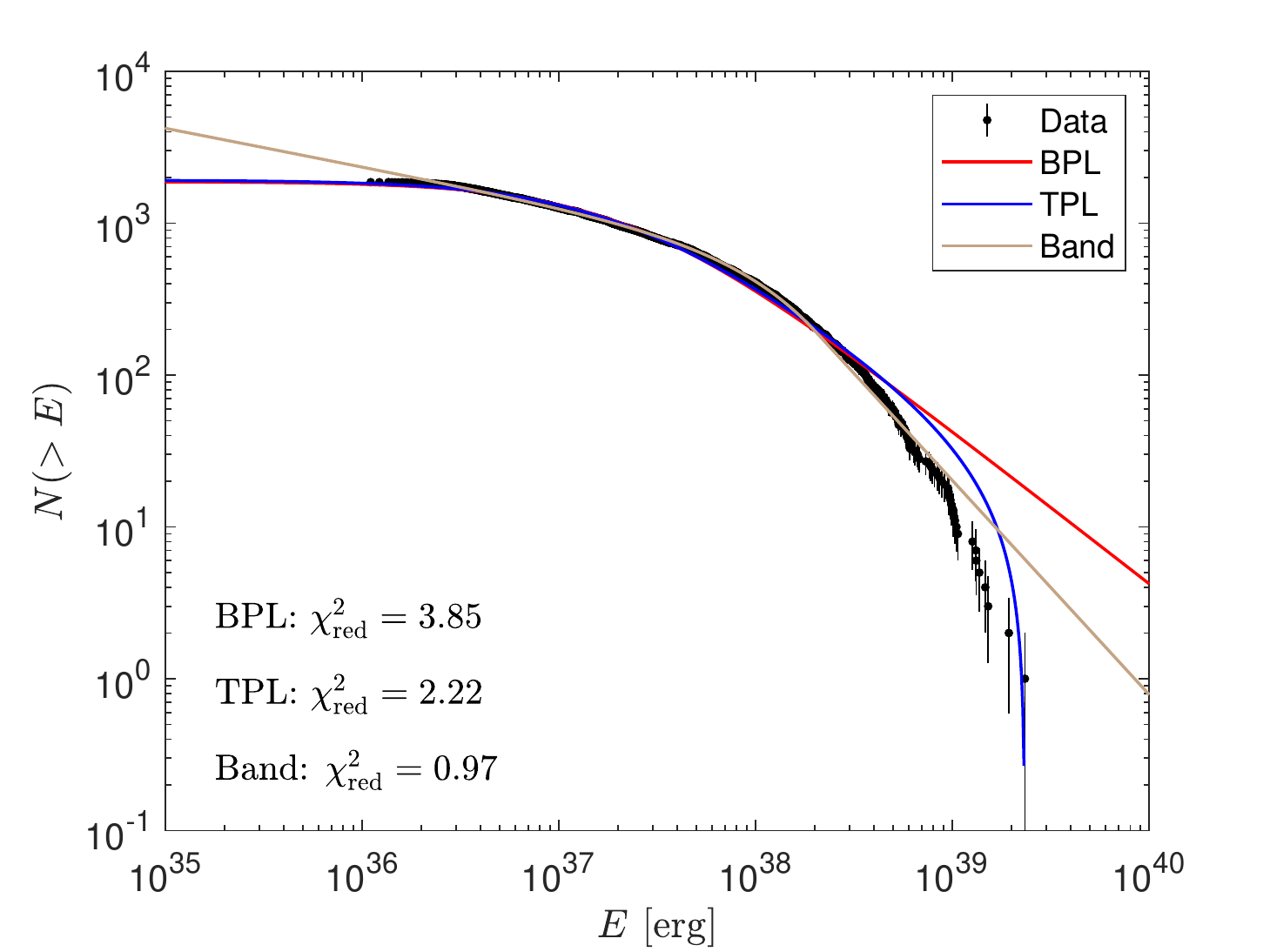}
	\includegraphics[width=0.48\textwidth]{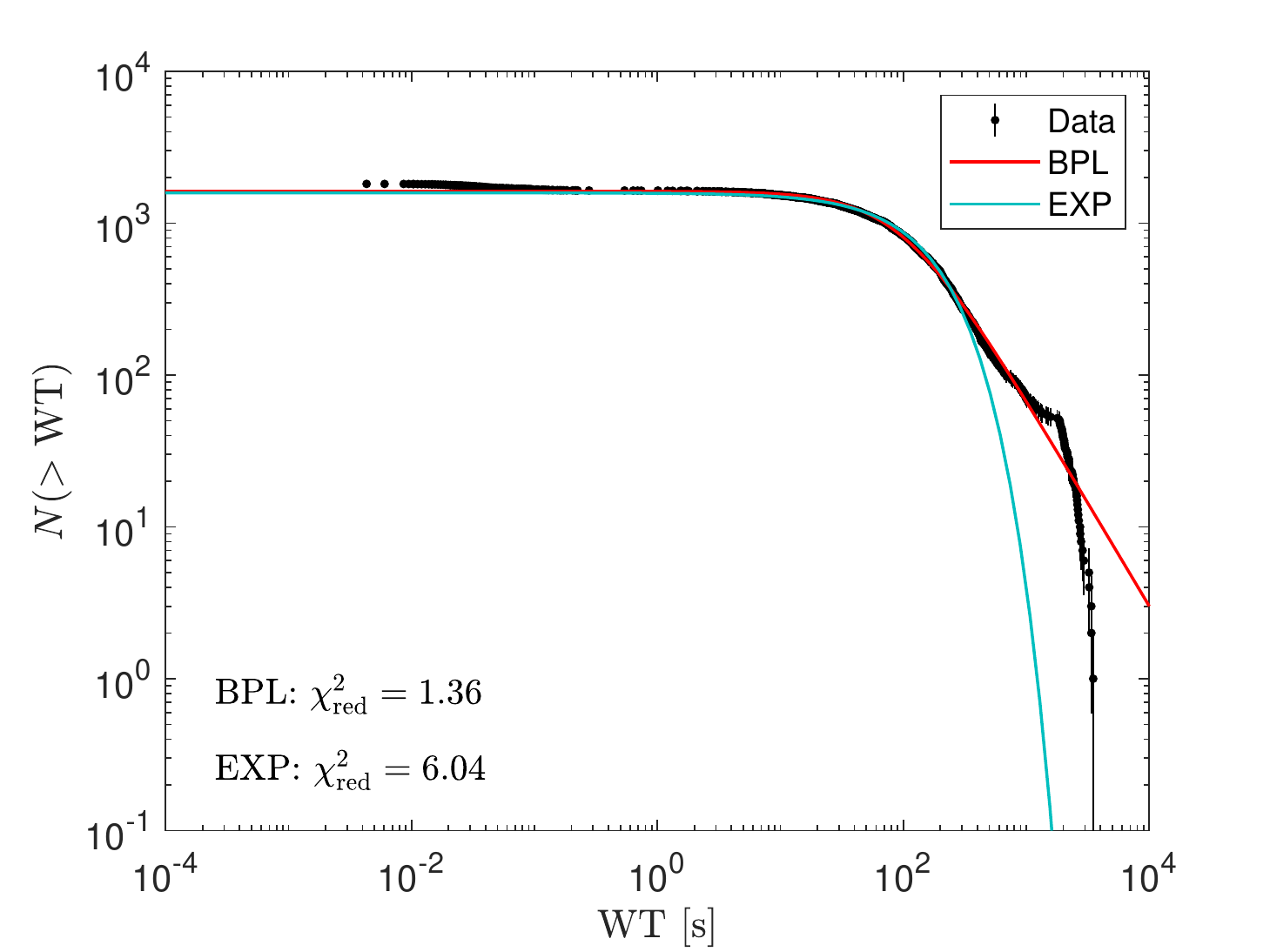}
    \caption{The CDFs of energy (left panel) and waiting time (right panel) for FRB 20201124A. In the left panel, the red, blue and light brown solid lines are the best-fitting curves to the BPL, TPL and Band models, respectively. In the right panel, the red and green solid lines are the best-fitting curves to the BPL and EXP models, respectively.}
    \label{fig:frb201124_full}
\end{figure*}

The fitting results for the energy distribution of FRB 20201124A are shown in the left panel of Figure \ref{fig:frb201124_full}, with the best-fitting parameters listed in Table \ref{tab:frb201124_full}. Similar to the fitting results of FRB 20121102A,  Both BPL and TPL models fit the low energy and middle energy parts well. In high energy part ($E\gtrsim 4\times10^{38} {\rm erg}$), however, both models fails to fit the data points. Especially, the BPL prediction significantly excesses the data points at the right most end. The best-fitting BPL parameters are $\beta = 1.01\pm0.01$ and $x_b = (2.36\pm0.03)\times 10^{37} {\rm erg}$, and the power-law index $\beta$ is similar to that of FRB 20121102A. The best-fitting TPL parameters are $\gamma = 1.85\pm0.01$ and $x_0 = (1.82\pm0.03) \times 10^{37} {\rm erg}$, and the power law index $\gamma$ is also similar to that of FRB 20121102A. The reduced chi-square values for BPL and TPL models are $\chi^2_{\rm red}= 3.85$ and $\chi^2_{\rm red}= 2.22$, respectively. As for the Band model, the best-fitting parameters are $\hat\alpha=-0.25\pm0.01$, $\hat\beta=-1.41\pm0.01$ and $x_c=(1.74\pm 0.01)\times{\rm 10^{38}erg}$. Although the overall reduced chi-square value for the Band model ($\chi^2_{\rm red}=0.97$) is smaller than that of BPL model and TPL model, the Band model fails to fit the data points at left-most and right-most ends.

The fitting results for the waiting time distribution of FRB 20201124A are shown in the right panel of Figure \ref{fig:frb201124_full}, with the best-fitting parameters listed in Table \ref{tab:frb201124_full}. As is seen, both BPL and EXP models fit the data points very well at middle part, but fail to fit the data points at left and right ends. At left end, both models are below the data points. At right end, the BPL prediction is above the data points, while the EXP prediction is below the data points. These features are similar with that of FRB 20121102A. A notable feature in the waiting time distribution of FRB 20201124A is that there is a sharp bump at ${\rm WT}\sim 2\times 10^{3}~{\rm s}$, which is absent in FRB 20121102A. The best-fitting BPL parameters are $\beta = 1.36\pm0.01$ and $x_b = 99.37\pm0.37 {\rm s}$, and the best-fitting EXP parameters are $\lambda = (5.98\pm0.03) \times 10^{-3} {\rm s}^{-1}$. From the reduced chi-square values, we can see that BPL model ($\chi^2_{\rm red}= 1.36$) is much better than EXP model ($\chi^2_{\rm red}= 6.04$), similar to the results of FRB 20121102A.

\begin{table*}
\centering
\caption{The best-fitting parameters of the BPL, TPL, Band and EXP models for FRB 20201124A.}\label{tab:frb201124_full}

{\begin{tabular}{lllll} 
\hline\hline 
\multicolumn{5}{c}{energy}\\
\hline
BPL & & $\beta=1.01\pm0.01$  & $x_b=(2.36\pm0.03)\times{\rm 10^{37}erg}$ & $\chi^2_{\rm red}=3.85$\\
TPL & & $\gamma=1.85\pm0.01$ & $x_0=(1.82\pm0.03)\times{\rm 10^{37}erg}$ & $\chi^2_{\rm red}=2.22$\\
Band & $\hat\alpha=-0.25\pm0.01$ & $\hat\beta=-1.41\pm0.01$ & $x_c=(1.74\pm 0.01)\times{\rm 10^{38}erg}$ & $\chi^2_{\rm red}=0.97$\\
\hline
\multicolumn{5}{c}{waiting time}\\
\hline
BPL & & $\beta=1.36\pm0.01$ & $x_b=99.37\pm0.37~{\rm s}$ & $\chi^2_{\rm red}=1.36$\\
EXP & & & $\lambda=(5.98\pm0.03)\times{\rm 10^{-3}s^{-1}}$ & $\chi^2_{\rm red}=6.04$\\
\hline 
\end{tabular}}
\end{table*}

To summarize, for the energy distributions of both samples, TPL model is better than BPL model; while for the waiting time distributions, BPL model is better than EXP model. However, none of these 3-parameter models can fit the data well in the whole range. Even the 4-parameter model (Band function) couldn't fit the whole range of the data points well. One reason may be that the statistical properties of the repeating FRBs are temporally evolving. This inspires us to bin the data and fit the models to each subsample separately.

\section{The statistics in different observing sessions}\label{sec:sub-sample}

In the FAST sample of FRB 20121102A we used in this paper, the energy distribution is found to be bimodal. The bimodal energy distribution is also time dependent, with more high energy bursts detected before MJD 58740 \citep{Li:2021hpl}. Dividing the data into two epochs according to MJD 58740, the high energy bursts of early stage have an inconsistent energy distribution with the late stage bursts \citep{Zhang:2021ztz}. In addition, it is found that the burst rate of FRB 20201124A observed in two observation stages differs significantly \citep{Zhang:2022rib}. These imply that the burst energy and burst rate may be temporally evolving. Therefore, in this section, we study the possible temporal evolution of energy and waiting time by dividing the full sample into different observing sessions.

Both FRB 20121102A and FRB 20201124A are observed by FAST in a period spanning about two months for each  \citep{Li:2021hpl,Xu:2021qdn}. However, the FAST observation is not continuous, with roughly 1 or 2 hours in each day. Therefore, bursts are gathered into the same subsample if the waiting time to the adjacent previous burst is smaller than 1 hr. This ensures that bursts in the same subsample are observed in the same observing session. To ensure the statistical significance, we only consider the subsamples with number of bursts $N\geq 30$. We find 22 subsamples for FRB 20121102A and 26 subsamples for FRB 20201124A, e.g., see Table \ref{tab:frb121102_binned} and Table \ref{tab:frb201124_binned} respectively. To distinguish different bursts from the same FRB source, we give each burst an identifier (ID). For example, the 1652 bursts from the repeating FRB 20121102A are labeled from 1 to 1652 according to the temporal order. The burst IDs in each subsample are listed in the second column in Table \ref{tab:frb121102_binned}, e.g., the first subsample includes the bursts labeled by ID = 1 to 87, which contains 87 bursts. Similar to the previous section, we use BPL and TPL models to fit the distribution of energy for each subsample, and use BPL and EXP models to fit the distribution of waiting time. We note that the Band model has one more parameter than the BPL model and TPL model. For the binned data, it is unnecessary to use the complex Band model to fit the energy distribution. In addition, the physical explanation for the Band model is still unclear. So we do not consider the Band model for the subsamples.

\begin{figure*}
\centering
	\includegraphics[width=0.48\textwidth]{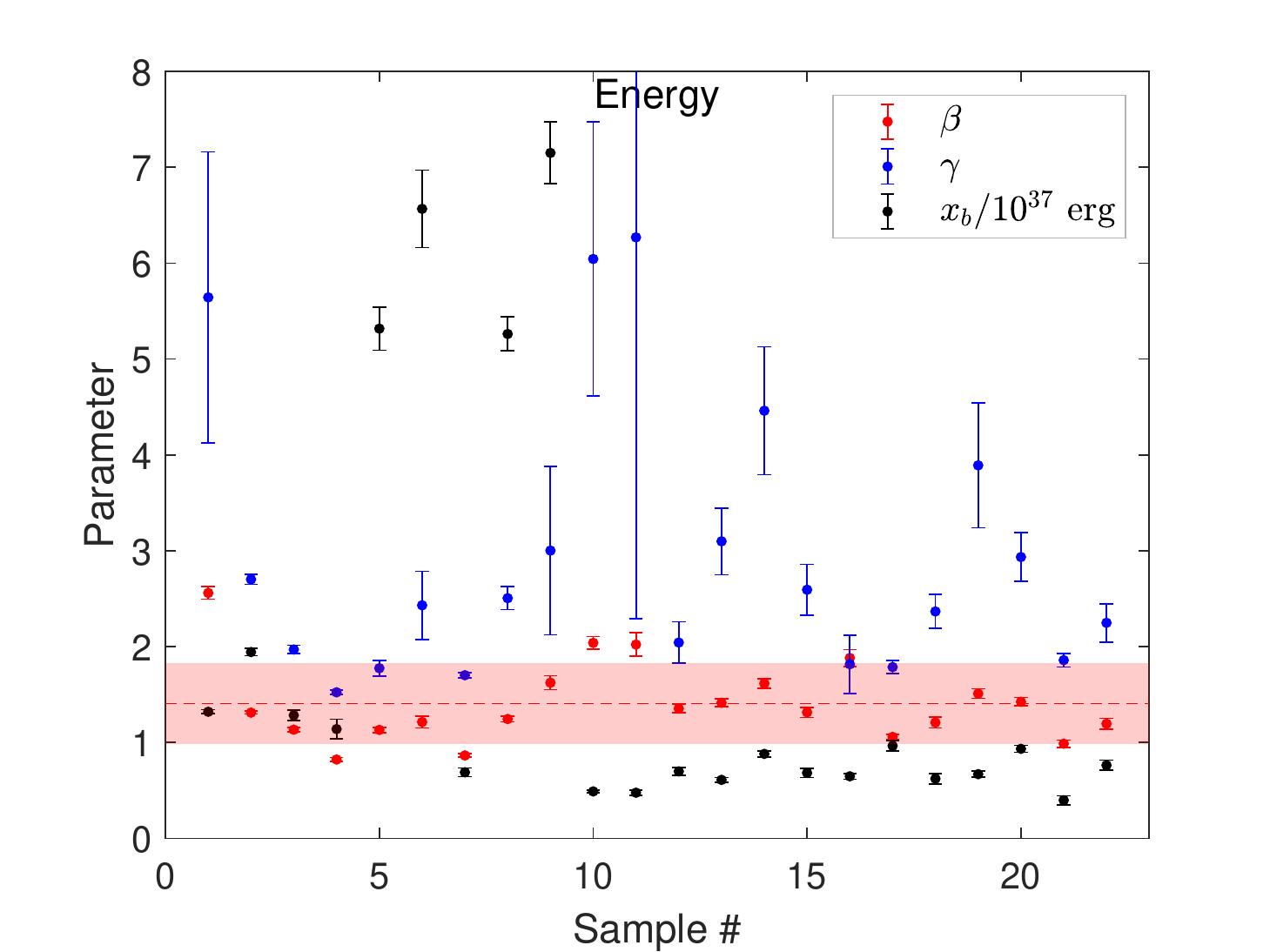}
    \includegraphics[width=0.48\textwidth]{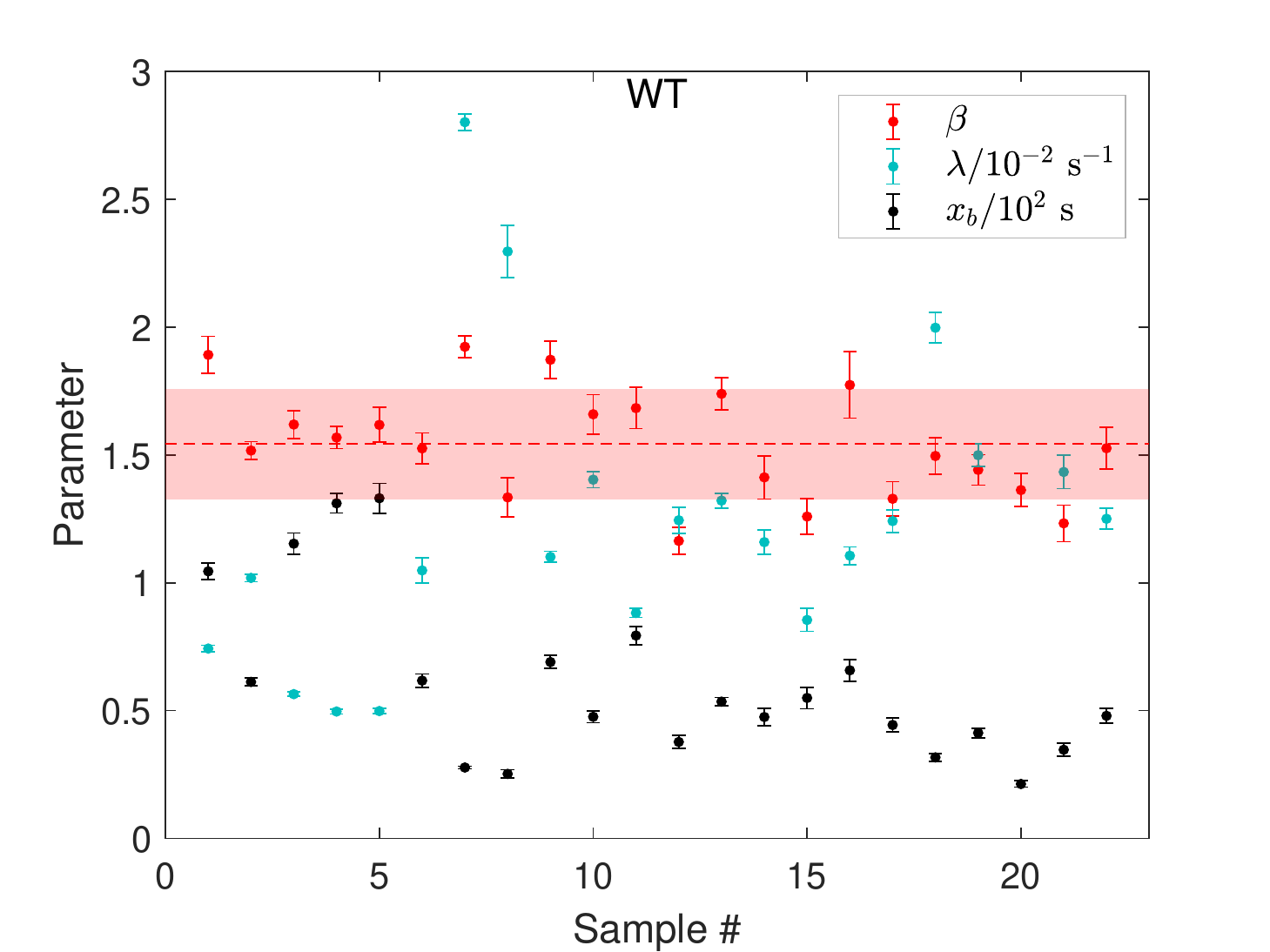}
    \caption{The best-fitting parameters for FRB 20121102A in different observing sessions. Left panel: the BPL parameters ($\beta$ and $x_b$) and TPL index ($\gamma$) for the energy distribution; right panel: the BPL parameters ($\beta$ and $x_b$) and EXP index ($\lambda$) for the WT distribution. The red-dashed line is the mean value of the BPL index ($\beta$) and the red-shaded region is the standard deviation. For energy $\bar{\beta}=1.41$, $\sigma_\beta=0.42$; for WT $\bar{\beta}=1.54$, $\sigma_\beta=0.22$.}
    \label{fig:frb121102_param}
\end{figure*}

The best-fitting parameters for the 22 subsamples of FRB 20121102A are summarized in Table \ref{tab:frb121102_binned}. The CDFs and the best-fitting curves of energy for all subsamples are shown in Figure \ref{fig:frb121102_energy} in Appendix \ref{sec:appendix}. For most subsamples, both BPL model and TPL model fit the distribution of energy equally well. The left panel of Figure \ref{fig:frb121102_param} depicts the temporal evolution of the power-law indices, i.e., the parameters $\beta$ and $\gamma$ as a function of subsample number. The best-fitting BPL index $\beta$ for each subsample is approximately invariant. The mean and standard deviation are $\bar{\beta}=1.41 \pm 0.42$, which are denoted by the red-dashed line and red-shaded region in the left panel of Figure \ref{fig:frb121102_param}, respectively. The best-fitting parameters $x_b$ in subsamples No. 5, 6, 8 and 9 are a few times larger than other subsamples. Note that $x_b$ is the median value of energy, which means that these subsamples have more energetic bursts than others. Compared to $\beta$, the best-fitting TPL indices $\gamma$ of energy distribution have a significant time dependence and larger uncertainties.

The CDFs and the best-fitting curves of waiting time for all subsamples of FRB 20121102A  are shown in \ref{fig:frb121102_wt} in Appendix \ref{sec:appendix}. For most subsamples, both BPL model and EXP model fit the distribution of waiting time equally well. The right panels of Figure \ref{fig:frb121102_param} show the temporal evolution of the parameters $\beta$ and $\lambda$. Similar to the energy, the best-fitting BPL index $\beta$ of waiting time for each subsample is also approximately invariant, with the mean and standard deviation $\bar{\beta} = 1.54\pm0.22$, but the median value parameter $x_b$ varies significantly. In addition, the best-fitting EXP parameter $\lambda$ has large fluctuation for different subsamples, indicating temporally evolving occurrence rate of bursts.

\begin{table*}
\centering
\setlength{\tabcolsep}{0.15cm}{
\caption{The best-fitting BPL, TPL and EXP parameters of energy and waiting time for FRB 20121102A in different observing sessions. The first column is the order of subsample, and the second column gives the burst IDs of the first and the last bursts in each subsample (see the text for details).}\label{tab:frb121102_binned}
{\begin{tabular}{lllllllllllll} 
\hline\hline 
 & & BPL(energy) & & & TPL(energy) & & & BPL(WT) & & & EXP(WT) &\\ 
\cline{3-13} 
No. & burst ID& $\beta$ & $x_b[\rm 10^{37}erg]$ & $\chi^2_{\rm red}$& $\gamma$ & $x_0[\rm 10^{37}erg]$ & $\chi^2_{\rm red}$& $\beta$ & $x_b[\rm s]$ & $\chi^2_{\rm red}$& $\lambda[\rm 10^{-2}s^{-1}]$ & $\chi^2_{\rm red}$ \\ 
\hline 
1 & $1-87$ & $2.56\pm0.06$ & $1.32\pm0.02$ & $0.18$ & $5.64\pm1.52$ & $4.67\pm1.92$ & $0.70$ & $1.89\pm0.07$ & $104.47\pm3.23$ & $0.47$ & $0.74\pm0.01$ & $0.28$ \\ 
2 & $88-208$ & $1.31\pm0.02$ & $1.94\pm0.04$ & $0.08$ & $2.70\pm0.05$ & $3.41\pm0.17$ & $0.06$ & $1.52\pm0.04$ & $61.22\pm1.60$ & $0.42$ & $1.02\pm0.01$ & $0.30$ \\ 
3 & $209-318$ & $1.13\pm0.02$ & $1.28\pm0.05$ & $0.20$ & $1.97\pm0.04$ & $1.15\pm0.10$ & $0.15$ & $1.62\pm0.05$ & $115.29\pm4.18$ & $0.75$ & $0.56\pm0.01$ & $0.35$ \\ 
4 & $319-409$ & $0.82\pm0.02$ & $1.14\pm0.10$ & $0.21$ & $1.52\pm0.02$ & $0.46\pm0.07$ & $0.16$ & $1.57\pm0.04$ & $131.08\pm3.82$ & $0.30$ & $0.50\pm0.01$ & $0.28$ \\ 
5 & $410-474$ & $1.13\pm0.03$ & $5.32\pm0.22$ & $0.12$ & $1.78\pm0.08$ & $3.80\pm0.67$ & $0.13$ & $1.62\pm0.07$ & $133.06\pm5.81$ & $0.38$ & $0.50\pm0.01$ & $0.20$ \\ 
6 & $475-510$ & $1.21\pm0.06$ & $6.57\pm0.40$ & $0.12$ & $2.43\pm0.36$ & $10.37\pm3.63$ & $0.17$ & $1.53\pm0.06$ & $61.74\pm2.60$ & $0.10$ & $1.05\pm0.05$ & $0.23$ \\ 
7 & $527-632$ & $0.87\pm0.02$ & $0.69\pm0.05$ & $0.17$ & $1.70\pm0.03$ & $0.45\pm0.05$ & $0.18$ & $1.92\pm0.04$ & $27.75\pm0.51$ & $0.25$ & $2.80\pm0.03$ & $0.16$ \\ 
8 & $633-713$ & $1.25\pm0.03$ & $5.26\pm0.18$ & $0.15$ & $2.51\pm0.12$ & $8.81\pm1.04$ & $0.12$ & $1.33\pm0.08$ & $25.26\pm1.60$ & $0.81$ & $2.30\pm0.10$ & $0.93$ \\ 
9 & $714-752$ & $1.62\pm0.07$ & $7.15\pm0.32$ & $0.15$ & $3.00\pm0.88$ & $16.46\pm8.65$ & $0.22$ & $1.87\pm0.07$ & $69.00\pm2.54$ & $0.12$ & $1.10\pm0.02$ & $0.07$ \\ 
10 & $782-839$ & $2.04\pm0.07$ & $0.49\pm0.01$ & $0.13$ & $6.04\pm1.43$ & $2.28\pm0.78$ & $0.17$ & $1.66\pm0.08$ & $47.58\pm2.30$ & $0.39$ & $1.40\pm0.03$ & $0.19$ \\ 
11 & $856-892$ & $2.02\pm0.12$ & $0.48\pm0.03$ & $0.19$ & $6.27\pm3.98$ & $2.32\pm2.11$ & $0.32$ & $1.68\pm0.08$ & $79.37\pm3.60$ & $0.14$ & $0.88\pm0.02$ & $0.06$ \\ 
12 & $893-944$ & $1.35\pm0.05$ & $0.70\pm0.04$ & $0.15$ & $2.04\pm0.22$ & $0.67\pm0.23$ & $0.24$ & $1.16\pm0.05$ & $37.74\pm2.52$ & $0.28$ & $1.24\pm0.05$ & $0.38$ \\ 
13 & $966-1018$ & $1.42\pm0.04$ & $0.61\pm0.02$ & $0.08$ & $3.10\pm0.35$ & $1.34\pm0.32$ & $0.15$ & $1.74\pm0.06$ & $53.51\pm1.67$ & $0.17$ & $1.32\pm0.03$ & $0.15$ \\ 
14 & $1019-1068$ & $1.62\pm0.05$ & $0.88\pm0.03$ & $0.08$ & $4.46\pm0.67$ & $3.28\pm0.82$ & $0.12$ & $1.41\pm0.08$ & $47.51\pm3.41$ & $0.48$ & $1.16\pm0.05$ & $0.39$ \\ 
15 & $1069-1109$ & $1.32\pm0.05$ & $0.68\pm0.05$ & $0.08$ & $2.59\pm0.26$ & $1.05\pm0.30$ & $0.13$ & $1.26\pm0.07$ & $54.96\pm4.22$ & $0.29$ & $0.85\pm0.04$ & $0.37$ \\ 
16 & $1110-1154$ & $1.88\pm0.09$ & $0.65\pm0.03$ & $0.20$ & $1.82\pm0.30$ & $0.62\pm0.25$ & $0.07$ & $1.77\pm0.13$ & $65.77\pm4.23$ & $0.44$ & $1.11\pm0.03$ & $0.23$ \\ 
17 & $1155-1208$ & $1.06\pm0.03$ & $0.97\pm0.05$ & $0.08$ & $1.79\pm0.07$ & $0.69\pm0.11$ & $0.10$ & $1.33\pm0.07$ & $44.39\pm2.73$ & $0.34$ & $1.24\pm0.04$ & $0.32$ \\ 
18 & $1209-1261$ & $1.21\pm0.06$ & $0.62\pm0.06$ & $0.22$ & $2.37\pm0.18$ & $0.84\pm0.20$ & $0.27$ & $1.50\pm0.07$ & $31.71\pm1.55$ & $0.29$ & $2.00\pm0.06$ & $0.25$ \\ 
19 & $1262-1321$ & $1.51\pm0.05$ & $0.67\pm0.03$ & $0.16$ & $3.89\pm0.65$ & $2.16\pm0.65$ & $0.23$ & $1.44\pm0.06$ & $41.29\pm1.87$ & $0.29$ & $1.50\pm0.04$ & $0.29$ \\ 
20 & $1342-1458$ & $1.43\pm0.04$ & $0.94\pm0.04$ & $0.38$ & $2.94\pm0.25$ & $1.76\pm0.35$ & $0.54$ & $1.36\pm0.07$ & $21.31\pm1.24$ & $1.43$ & $2.58\pm0.07$ & $1.09$ \\ 
21 & $1459-1522$ & $0.99\pm0.04$ & $0.40\pm0.05$ & $0.19$ & $1.86\pm0.07$ & $0.31\pm0.07$ & $0.23$ & $1.23\pm0.07$ & $34.69\pm2.61$ & $0.61$ & $1.43\pm0.06$ & $0.67$ \\ 
22 & $1523-1575$ & $1.20\pm0.06$ & $0.76\pm0.05$ & $0.21$ & $2.25\pm0.20$ & $0.94\pm0.25$ & $0.31$ & $1.53\pm0.08$ & $47.97\pm2.87$ & $0.44$ & $1.25\pm0.04$ & $0.31$ \\ 
\hline 
\end{tabular}} 
}
\end{table*}

\begin{figure*}
\centering
	\includegraphics[width=0.48\textwidth]{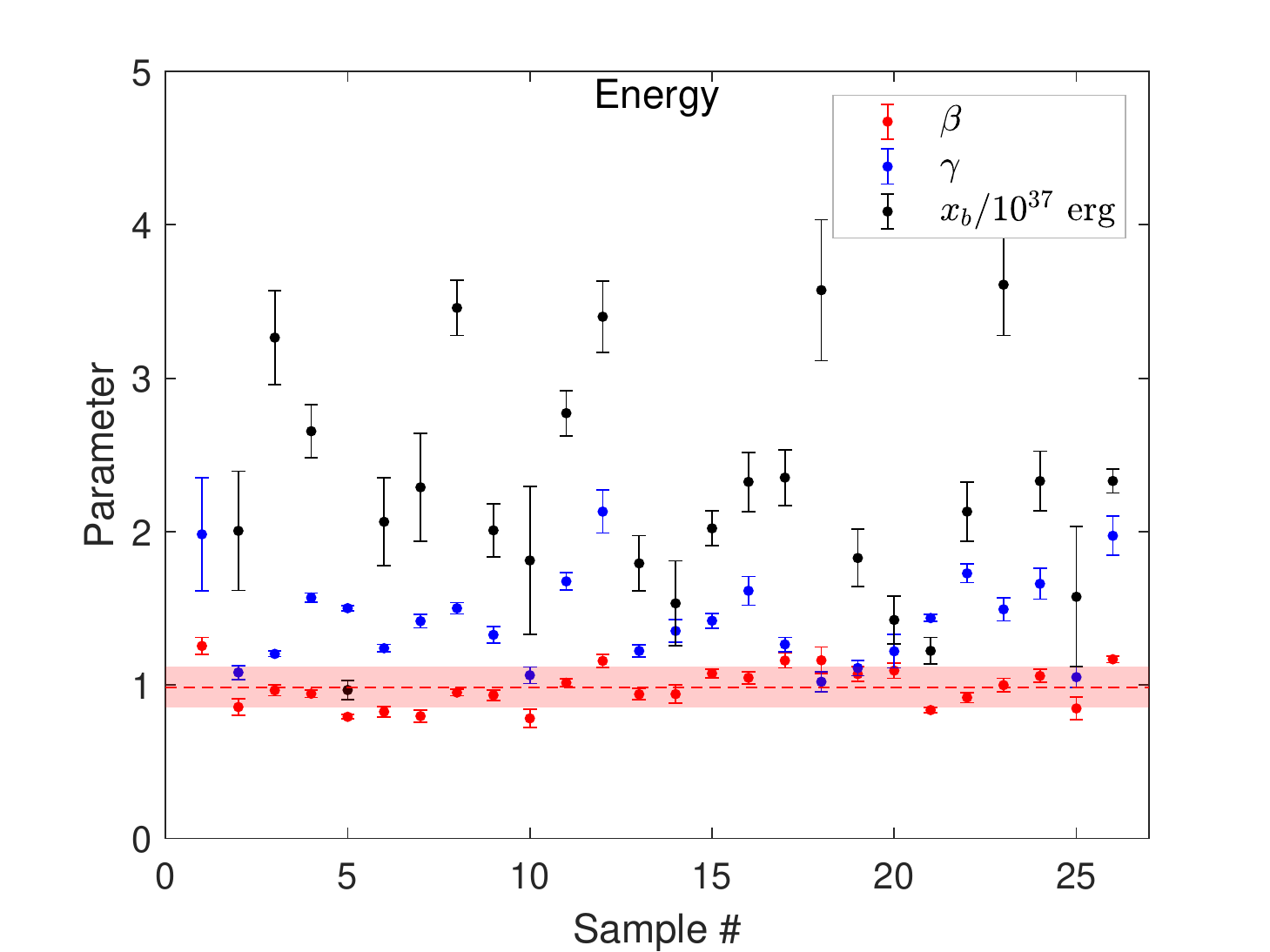}
 \includegraphics[width=0.48\textwidth]{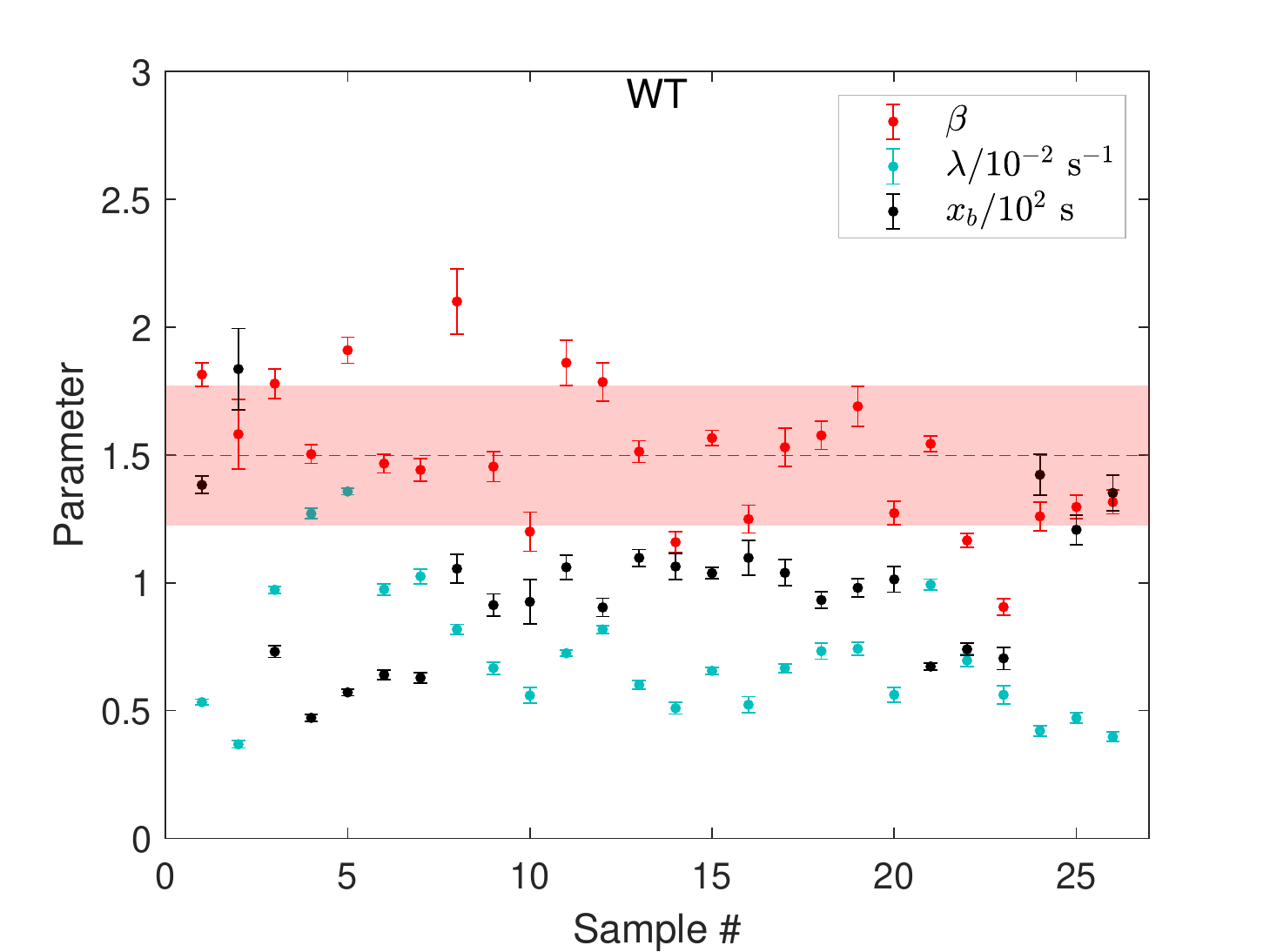}
    \caption{The best-fitting parameters for FRB 20201124A in different observing sessions. Left panel: the BPL parameters ($\beta$ and $x_b$) and TPL index ($\gamma$) for the energy distribution; right panel: the BPL parameters ($\beta$ and $x_b$) and EXP index ($\lambda$) for the WT distribution. The red-dashed line is the mean value of the index ($\beta$) and the red-shaded region is the standard deviation. For energy $\bar{\beta}=0.98$, $\sigma_\beta=0.13$; for WT $\bar{\beta}=1.50$, $\sigma_\beta=0.28$.}
    \label{fig:frb201124_param}
\end{figure*}

The best-fitting parameters for the 26 subsamples of FRB 20201124A are summarized in Table \ref{tab:frb201124_binned}. The CDFs and best-fitting curves of energy and waiting time for all subsamples are shown in Figure \ref{fig:frb201124_energy} and \ref{fig:frb201124_wt} in Appendix \ref{sec:appendix}, respectively. For most subsamples, both BPL model and TPL model fit the energy distribution equally well, and both BPL model and EXP model fit the waiting time distribution equally well. The temporal evolution of the best-fitting parameters $\beta$, $\gamma$ and $\lambda$ are shown in Figure \ref{fig:frb201124_param}. The temporal evolution of parameters are very similar to that of FRB 20121102A. For example, the BPL index $\beta$ is approximately invariant for both energy and waiting time, with the mean value and standard deviation $\bar{\beta}=0.98 \pm 0.13$ for energy distribution and $\bar{\beta}=1.50 \pm 0.28$ for waiting time distribution. The median value parameters in BPL model $x_b$ for both energy and waiting time vary significantly. Large fluctuations exist in the time evolution of the best-fitting TPL parameters $\gamma$ for energy distribution and EXP parameters $\lambda$ for waiting time distribution.

\begin{table*}
\centering
\setlength{\tabcolsep}{0.15cm}{
\caption{The best-fitting BPL, TPL and EXP parameters of energy and waiting time for FRB 20201124A in different observing sessions. The first column is the order of subsample, and the second column gives the burst IDs of the first and the last bursts in each subsample (see the text for details).}\label{tab:frb201124_binned}
{\begin{tabular}{lllllllllllll} 
\hline\hline 
 & & BPL(energy) & & & TPL(energy) & & & BPL(WT) & & & EXP(WT) &\\ 
\cline{3-13} 
No. & burst ID& $\beta$ & $x_b[\rm 10^{37}erg]$ & $\chi^2_{\rm red}$& $\gamma$ & $x_0[\rm 10^{37}erg]$ & $\chi^2_{\rm red}$& $\beta$ & $x_b[\rm s]$ & $\chi^2_{\rm red}$& $\lambda[\rm 10^{-2}s^{-1}]$ & $\chi^2_{\rm red}$ \\ 
\hline 
1 & $95-133$ & $1.25\pm0.06$ & $8.13\pm0.45$ & $0.12$ & $1.98\pm0.37$ & $8.61\pm4.23$ & $0.22$ & $1.81\pm0.05$ & $138.26\pm3.43$ & $0.05$ & $0.53\pm0.01$ & $0.07$ \\ 
2 & $134-166$ & $0.86\pm0.06$ & $2.01\pm0.39$ & $0.17$ & $1.08\pm0.05$ & $0.02\pm0.09$ & $0.11$ & $1.58\pm0.14$ & $183.59\pm15.86$ & $0.29$ & $0.37\pm0.01$ & $0.16$ \\ 
3 & $167-248$ & $0.97\pm0.04$ & $3.26\pm0.31$ & $0.39$ & $1.20\pm0.02$ & $0.27\pm0.07$ & $0.07$ & $1.78\pm0.06$ & $73.06\pm2.26$ & $0.37$ & $0.97\pm0.01$ & $0.14$ \\ 
4 & $249-350$ & $0.94\pm0.02$ & $2.65\pm0.17$ & $0.25$ & $1.57\pm0.03$ & $1.10\pm0.14$ & $0.16$ & $1.50\pm0.04$ & $47.17\pm1.46$ & $0.34$ & $1.27\pm0.02$ & $0.25$ \\ 
5 & $351-458$ & $0.79\pm0.01$ & $0.97\pm0.06$ & $0.12$ & $1.50\pm0.02$ & $0.37\pm0.04$ & $0.10$ & $1.91\pm0.05$ & $57.13\pm1.33$ & $0.41$ & $1.36\pm0.01$ & $0.13$ \\ 
6 & $459-531$ & $0.83\pm0.03$ & $2.06\pm0.29$ & $0.36$ & $1.24\pm0.02$ & $0.20\pm0.07$ & $0.16$ & $1.47\pm0.04$ & $64.04\pm1.79$ & $0.16$ & $0.97\pm0.02$ & $0.24$ \\ 
7 & $532-586$ & $0.80\pm0.04$ & $2.29\pm0.35$ & $0.21$ & $1.42\pm0.04$ & $0.55\pm0.18$ & $0.20$ & $1.44\pm0.04$ & $62.84\pm2.11$ & $0.11$ & $1.03\pm0.03$ & $0.22$ \\ 
8 & $587-647$ & $0.95\pm0.02$ & $3.46\pm0.18$ & $0.07$ & $1.50\pm0.04$ & $1.30\pm0.20$ & $0.07$ & $2.10\pm0.13$ & $105.50\pm5.51$ & $0.61$ & $0.82\pm0.02$ & $0.29$ \\ 
9 & $648-694$ & $0.93\pm0.03$ & $2.01\pm0.17$ & $0.10$ & $1.33\pm0.05$ & $0.37\pm0.14$ & $0.11$ & $1.45\pm0.06$ & $91.29\pm4.40$ & $0.17$ & $0.67\pm0.02$ & $0.24$ \\ 
10 & $695-726$ & $0.78\pm0.06$ & $1.81\pm0.48$ & $0.20$ & $1.06\pm0.05$ & $0.00\pm0.12$ & $0.12$ & $1.20\pm0.08$ & $92.57\pm8.66$ & $0.16$ & $0.56\pm0.03$ & $0.25$ \\ 
11 & $727-782$ & $1.01\pm0.03$ & $2.77\pm0.15$ & $0.08$ & $1.68\pm0.06$ & $1.59\pm0.27$ & $0.09$ & $1.86\pm0.09$ & $106.07\pm4.84$ & $0.34$ & $0.72\pm0.01$ & $0.11$ \\ 
12 & $783-846$ & $1.16\pm0.04$ & $3.40\pm0.23$ & $0.27$ & $2.13\pm0.14$ & $3.99\pm0.84$ & $0.26$ & $1.79\pm0.08$ & $90.37\pm3.57$ & $0.36$ & $0.82\pm0.02$ & $0.17$ \\ 
13 & $847-893$ & $0.94\pm0.04$ & $1.79\pm0.18$ & $0.13$ & $1.22\pm0.04$ & $0.14\pm0.08$ & $0.09$ & $1.51\pm0.04$ & $109.75\pm3.27$ & $0.08$ & $0.60\pm0.02$ & $0.17$ \\ 
14 & $894-943$ & $0.94\pm0.06$ & $1.53\pm0.28$ & $0.35$ & $1.35\pm0.07$ & $0.28\pm0.17$ & $0.29$ & $1.16\pm0.04$ & $106.36\pm5.11$ & $0.13$ & $0.51\pm0.02$ & $0.38$ \\ 
15 & $944-1008$ & $1.08\pm0.03$ & $2.02\pm0.11$ & $0.12$ & $1.42\pm0.05$ & $0.52\pm0.13$ & $0.10$ & $1.57\pm0.03$ & $103.75\pm2.21$ & $0.07$ & $0.66\pm0.01$ & $0.18$ \\ 
16 & $1009-1048$ & $1.05\pm0.04$ & $2.32\pm0.19$ & $0.10$ & $1.61\pm0.09$ & $1.24\pm0.35$ & $0.11$ & $1.25\pm0.05$ & $109.75\pm6.75$ & $0.14$ & $0.52\pm0.03$ & $0.42$ \\ 
17 & $1049-1110$ & $1.16\pm0.05$ & $2.35\pm0.18$ & $0.28$ & $1.26\pm0.05$ & $0.36\pm0.12$ & $0.10$ & $1.53\pm0.08$ & $103.90\pm5.09$ & $0.30$ & $0.67\pm0.02$ & $0.24$ \\ 
18 & $1111-1147$ & $1.16\pm0.09$ & $3.57\pm0.46$ & $0.27$ & $1.02\pm0.06$ & $0.00\pm0.19$ & $0.12$ & $1.58\pm0.06$ & $93.24\pm3.31$ & $0.08$ & $0.73\pm0.03$ & $0.22$ \\ 
19 & $1148-1197$ & $1.07\pm0.05$ & $1.83\pm0.19$ & $0.24$ & $1.11\pm0.05$ & $0.11\pm0.10$ & $0.13$ & $1.69\pm0.08$ & $98.04\pm3.67$ & $0.20$ & $0.74\pm0.03$ & $0.30$ \\ 
20 & $1198-1240$ & $1.09\pm0.05$ & $1.42\pm0.16$ & $0.18$ & $1.22\pm0.11$ & $0.18\pm0.20$ & $0.26$ & $1.27\pm0.05$ & $101.31\pm5.02$ & $0.11$ & $0.56\pm0.03$ & $0.35$ \\ 
21 & $1241-1317$ & $0.84\pm0.02$ & $1.22\pm0.09$ & $0.09$ & $1.44\pm0.02$ & $0.33\pm0.06$ & $0.08$ & $1.54\pm0.03$ & $67.25\pm1.41$ & $0.11$ & $0.99\pm0.02$ & $0.26$ \\ 
22 & $1355-1417$ & $0.92\pm0.03$ & $2.13\pm0.19$ & $0.17$ & $1.73\pm0.06$ & $1.44\pm0.27$ & $0.18$ & $1.17\pm0.03$ & $74.00\pm2.40$ & $0.09$ & $0.70\pm0.02$ & $0.38$ \\ 
23 & $1565-1608$ & $1.00\pm0.04$ & $3.61\pm0.33$ & $0.13$ & $1.49\pm0.08$ & $1.32\pm0.40$ & $0.12$ & $0.91\pm0.03$ & $70.45\pm4.39$ & $0.11$ & $0.56\pm0.04$ & $0.53$ \\ 
24 & $1635-1673$ & $1.06\pm0.04$ & $2.33\pm0.20$ & $0.09$ & $1.66\pm0.10$ & $1.32\pm0.38$ & $0.11$ & $1.26\pm0.06$ & $142.24\pm8.08$ & $0.11$ & $0.42\pm0.02$ & $0.29$ \\ 
25 & $1674-1716$ & $0.85\pm0.07$ & $1.58\pm0.46$ & $0.44$ & $1.05\pm0.07$ & $0.00\pm0.15$ & $0.22$ & $1.30\pm0.05$ & $120.72\pm5.80$ & $0.10$ & $0.47\pm0.02$ & $0.29$ \\ 
26 & $1717-1757$ & $1.17\pm0.02$ & $2.33\pm0.08$ & $0.02$ & $1.97\pm0.13$ & $2.22\pm0.48$ & $0.07$ & $1.32\pm0.05$ & $135.13\pm6.93$ & $0.11$ & $0.40\pm0.02$ & $0.29$ \\ 
\hline 
\end{tabular}} 
}
\end{table*}

In summary, for the energy distribution, the BPL parameters $\beta$ of both FRBs are approximately invariant, $\bar{\beta}=1.41 \pm 0.42$ for FRB 20121101A and $\bar{\beta}=0.98 \pm 0.13$ for FRB 20201124A. But the parameters $x_b$ vary significantly for both FRBs, implying the temporal evolution of burst energy. For the waiting time distribution, the BPL parameters $\beta$ of both FRBs are also approximately invariant, $\bar{\beta} = 1.54\pm0.22$  for FRB 20121101A and $\bar{\beta}=1.50 \pm 0.28$ for FRB 20201124A. But the parameters $x_b$ also vary significantly for both FRBs. In addition, the EXP parameters $\lambda$ for the waiting time distribution for both FRBs show significant variance, implying the temporal evolution of average burst rate.

\section{Discussions and Conclusions}\label{sec:conclusion}

In this paper, we investigate the statistical properties of energy and waiting time of two extremely active repeating FRBs observed by the FAST telescope, i.e. FRB 20121102A and FRB 20201124A. The distribution of energy is fitted by BPL, TPL and Band models, and the distribution of waiting time is fitted by BPL and EXP models for the full samples. It is found that no single model can fit the distribution of energy or waiting time well in the whole range. We therefore divide the full samples into several subsamples according to the observing sessions, to study the possible temporal evolution of statistical properties. It is found that the energy distribution of each subsample can be well fitted by BPL model or TPL model, and the waiting time distribution of each subsample can be well fitted by BPL model or EXP model. For the best-fitting BPL index $\beta$ for both energy and waiting time, the values are very close to each other for different subsamples, but the best-fitting median value parameters $x_b$ in BPL model varies significantly. For other parameters, including the TPL index $\gamma$ and EXP parameter $\lambda$, there is large fluctuation for different subsamples. These features imply that there may be a common emission mechanism for the repeating FRBs, but the burst energy and burst rate are temporally evolving.

The distribution of energy for each subsample is well fitted by BPL model. BPL is a smoothly connected peicewise power law, which has a flat tail at low-energy end ($x\ll x_b$), and reduces to simple power law at high-energy end ($x\gg x_b$). The flat tail at low-energy end implies that there is a lack of low-energy bursts, which can be explained by the observational incompleteness, since some dim bursts with fluence below or near the detection threshold couldn't be observed. To avoid the incompleteness, a common method is to discard the low-energy bursts below a certain threshold energy, and fit the high-energy bursts using the simple power-law model. Since the BPL model reduces to the simple power-law model in high-energy limit ($x\gg x_b$), the BPL index $\beta$ is expected to be close to the simple power-law index if fitting a simple power law model to high-energy bursts only.

Based on the samples of 1652 bursts from FRB 20121102A, and 1863 bursts from FRB 20201124A, our results show that the statistical properties of these two repeating FRBs have some similarity. First, the BPL index ($\beta$) of energy is almost temporally invariant. The average $\beta$ value of FRB 20121102A ($\bar{\beta}=1.41\pm 0.42$) is consistent with that of FRB 20201124A ($\bar{\beta}=0.98\pm 0.13$) within $1\sigma$ uncertainty. Second, the BPL index of waiting time is also temporally invariant, and the average BPL indices of these two FRBs are approximately equal, i.e. $\bar{\beta} = 1.54\pm 0.22$ for FRB 20121102A, and $\bar{\beta}=1.50\pm 0.28$ for FRB 20201124A. Third, the median value parameters ($x_b$) of both energy and waiting time show significant variation for both FRBs. Fourth, the TPL index ($\gamma$) of energy (especially for FRB 20121102A) shows strong variation, with $\bar\gamma=2.94\pm 1.43$ and $\bar\gamma=1.43\pm 0.30$ for FRB 20121102A and FRB 20201124A, respectively. Finally, the waiting time of each subsample can be well fitted by the EXP model. The mean values of $\lambda$ (in unit of $10^{-2}~{\rm s}^{-1}$) for FRB 20121102A and FRB 20201124A are $\bar\lambda=1.30\pm 0.63$ and $\bar\lambda=0.72\pm 0.25$, respectively. The $\lambda$ values for both FRBs show strong variation, especially for FRB 20121102A, implying temporally evolving occurrence rate. The most active episode of FRB 20121102A is the subsample No.7, with the average occurrence rate $\lambda=(2.80\pm 0.03)\times 10^{-2}~{\rm s}^{-1}$, equivalently $\lambda\approx 100~{\rm hr}^{-1}$. For FRB 20201124A, the most active episode is the subsample No.5, with the average occurrence rate $\lambda=(1.36\pm 0.01)\times 10^{-2}~{\rm s}^{-1}$, equivalently $\lambda\approx 49~{\rm hr}^{-1}$, which is half of FRB 20121102A.

The statistics of FRB 20121102A and FRB 20201124A were studied by \citet{Li:2021hpl} and \citet{Xu:2021qdn}, respectively. The probability density function (PDF) instead of the CDF of energy and waiting time was used in \citet{Li:2021hpl}. They used a log-normal distribution plus a generalized Cauchy function (four parameters in total) to fit the energy distribution, and used the log-normal function to fit each peak of the waiting time distribution. \citet{Xu:2021qdn} used a broken power law (four parameters) to fit the CDF of energy, and used the superposition of three log-normal distribution to fit the PDF of waiting time. Note that the PDF method depends on the binning of data points, so we use CDF rather than DPF in our paper. Moreover, the models we considered here are different from that of \citet{Li:2021hpl} and \citet{Xu:2021qdn}, with less parameters in our models than theirs. Additionally, the temporally evolving statistics have not been done in the previous two papers.

The statistical properties of the repeating FRB 20201124A were also studied using a sample of 881 bursts observed by FAST during an extremely active episode in the end of September 2021 \citep{Zhang:2022rib}, which was observed three months later than the sample used in our paper. The waiting time showed a double-peak distribution, which can be modeled with the superposition of two log-normal functions peaking at 51.22 ms and  10.05 s, respectively. The distribution of energy can be well fitted by the Band model, with the best-fitting parameters $\hat\alpha = -0.22\pm 0.01$, $\hat\beta = -3.27\pm 0.34$ and $x_c=(1.1\pm 0.2) \times 10^{39}~{\rm erg}$. However, the energy distributions of the two samples investigated in our paper couldn't be well fitted by the Band model. This implies that the Band model is not universal. In addition, \citet{Zhang:2022rib} found that the occurrence rate of FRB 20201124A initially increased exponentially, but the source activity suddenly stopped after the fourth day. The peak occurrence rate happened in the four day, with $\lambda=542~{\rm hr}^{-1}$, which is much higher than all other repeating FRBs currently known, including FRB 20201124A itself previously reported. This further confirms that the activity of repeating FRBs is temporally evolving.

\section*{Acknowledgements}

This work has been supported by the National Natural Science Fund of China (Grant Nos. 12005184, 12275034 and 12147102).

\section*{Data Availability}

The data of FRB 20121102A is available in \citet{Li:2021hpl}, and the data of FRB 20201124A is available in \citet{Xu:2021qdn}.



\bibliographystyle{mnras}
\bibliography{reference} 



\appendix

\section{extra figures}\label{sec:appendix}

\begin{figure*}
	\includegraphics[width=1.0\textwidth]{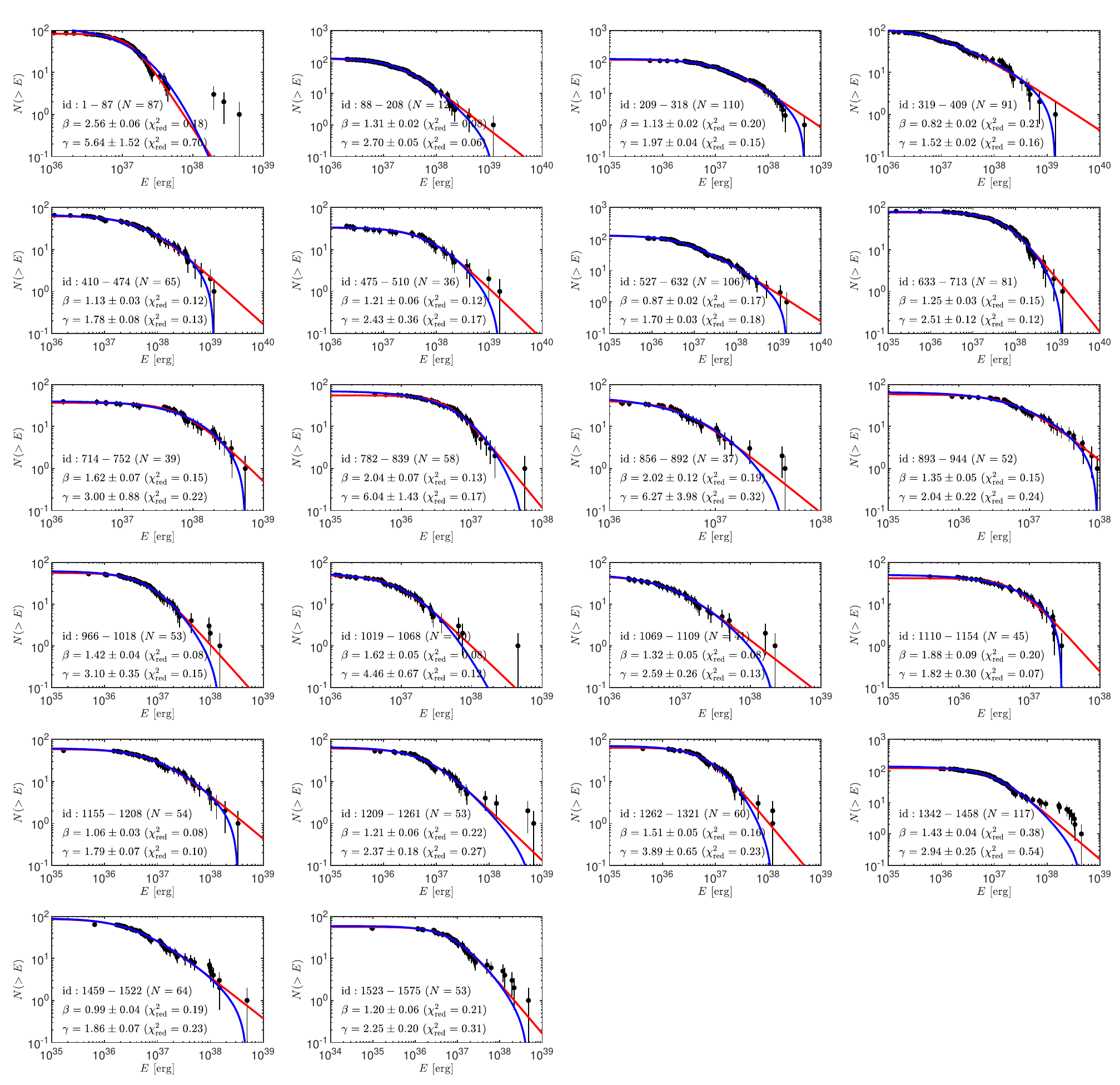}
    \caption{The CDFs of energy for FRB 20121102A in different observing sessions. The red and blue solid lines are the best-fitting curves to the BPL model and TPL model, respectively.}
    \label{fig:frb121102_energy}
\end{figure*}

\begin{figure*}
	\includegraphics[width=1.0\textwidth]{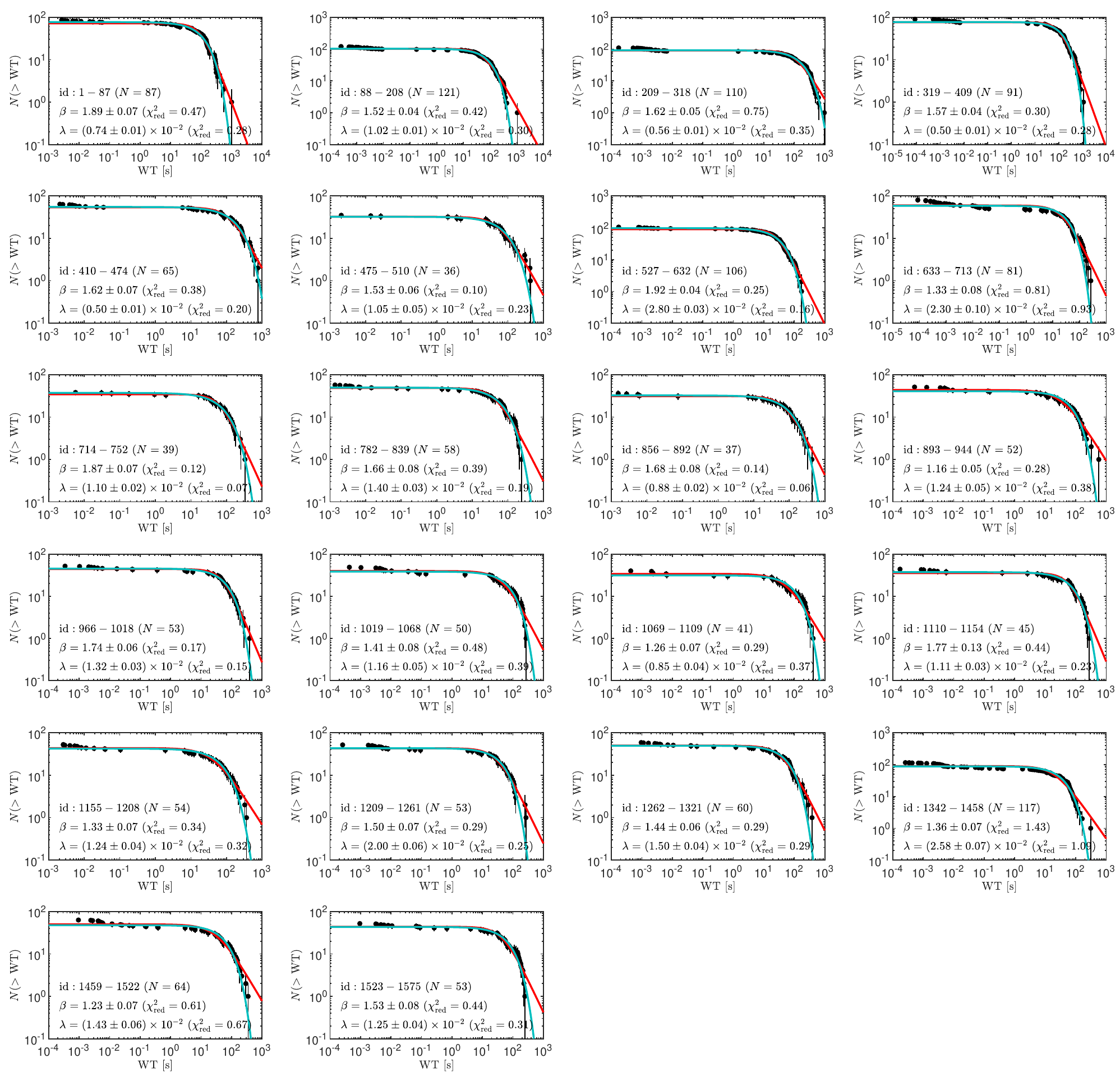}
    \caption{The CDFs of waiting time for FRB 20121102A in different observing sessions. The red and green solid lines are the best-fitting curves to the BPL model and EXP model, respectively.}
    \label{fig:frb121102_wt}
\end{figure*}

\begin{figure*}
	\includegraphics[width=1.0\textwidth]{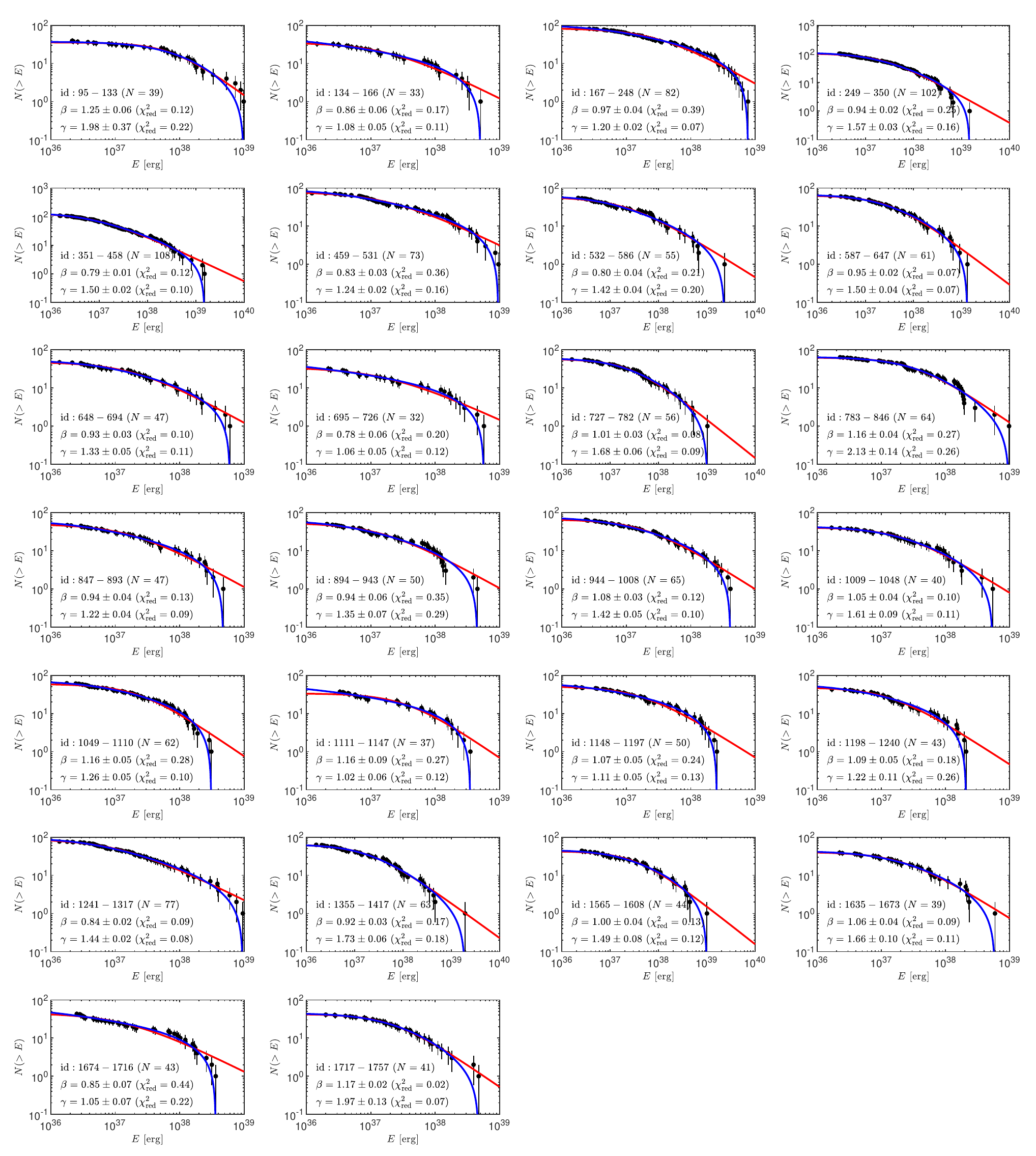}
    \caption{The CDFs of energy for FRB 20201124A in different observing sessions. The red and blue solid lines are the best-fitting curves to the BPL model and TPL model, respectively.}
    \label{fig:frb201124_energy}
\end{figure*}

\begin{figure*}
	\includegraphics[width=1.0\textwidth]{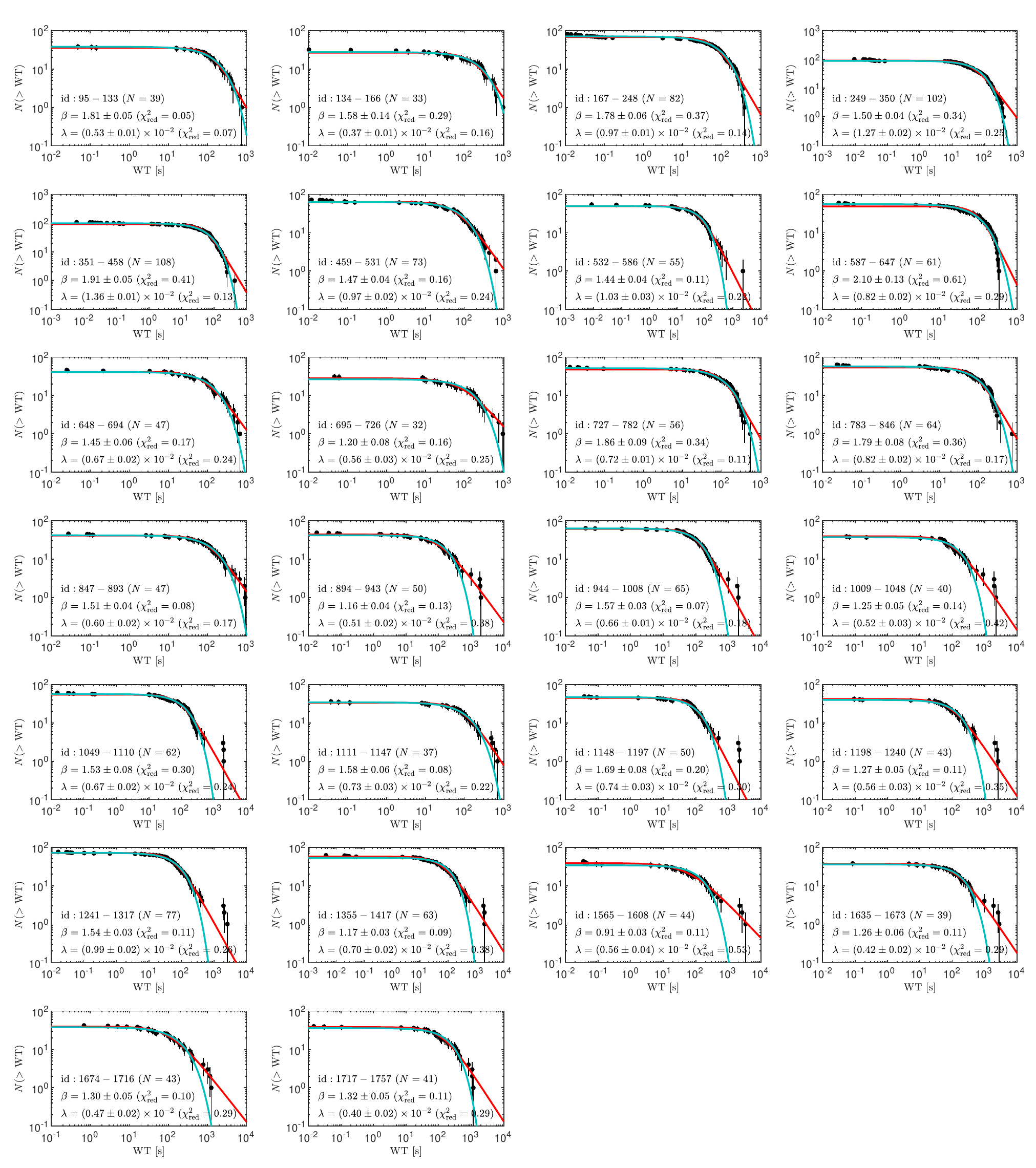}
    \caption{The CDFs of waiting time for FRB 20201124A in different observing sessions. The red and green solid lines are the best-fitting curves to the BPL model and EXP models, respectively.}
    \label{fig:frb201124_wt}
\end{figure*}

As was discussed in Section \ref{sec:sub-sample}, we studied the CDFs of energy and waiting time in different observing sessions. The best-fitting results for the energy and waiting time of FRB 20121102A are shown in Figure \ref{fig:frb121102_energy} and Figure \ref{fig:frb121102_wt}, respectively. The best-fitting results for the energy and waiting time of FRB 20201124A are shown in Figure \ref{fig:frb201124_energy} and Figure \ref{fig:frb201124_wt}, respectively.

\bsp	
\label{lastpage}
\end{document}